\documentstyle{elsart}
\begin{document}
\begin{frontmatter}


\title{ On a New Non--Geometric Element in Gravity}

\author{D. V. Ahluwalia}
\footnote{E-mail: av@p25hp.lanl.gov}
\address{P--25 Subatomic Physics Group,  Mail Stop H--846
\\ Los Alamos National Laboratory,
Los Alamos, NM 87545 USA \\
and\\
Global Power Division, ANSER Inc.,\\
1215 Jefferson Davis Highway\\
Arlington, VA 22202, USA}

\begin{abstract}
In this essay a generalized notion of flavor--oscillation clocks 
is introduced. The generalization contains the element that various
superimposed mass eigenstates  may have 
different relative orientation of the component of their spin 
with respect to the rotational axis of the the gravitational source.
It is found that these quantum mechanical clocks do not always 
redshift identically when moved from the gravitational
environment of a non--rotating source to the field of a rotating
source. 
The non--geometric contributions to the redshifts
may be interpreted as  quantum mechanically induced fluctuations
over  a
geometric structure of space--time.
\end{abstract}
\end{frontmatter}

Empirically observed equality of the inertial and gravitational masses 
leads to the
theory of general relativity \cite{SW}. This theory of gravity
lends itself to a
geometric interpretation.
In the framework of this theory,
all clocks, independent of
their workings, redshift in exactly the same manner for a given source.
It is further assumed that when these clocks move from the 
gravitational environment
of one source to another, 
identically running clocks run identically.

The primary purpose of this essay is to introduce the notion
of the {\em generalized flavor--oscillation
clocks}, and study their evolution
in a weak gravitational environment of a {\em rotating}
source. In the process we will uncover an inherently
non-geometric aspect in gravity.
Specifically, we will construct two 
intrinsically 
quantum mechanical clocks that
do not redshift identically when introduced in the 
gravitational environment of a {\em rotating} source.

It is a direct consequence of (a) the  1939 paper of Wigner \cite{EPW1939}, 
and
(b) the fact that {\em locally} 
space--time carries Poincar\'e symmetries; that a
general quantum test particle,
for  evolution over ``sufficiently small''
space--time distances,\footnote{For a global evolution one may need
to consider test particles that are characterized by Casimir
invariants associated with  global space--time symmetries 
of the relevant  gravitational source {\em a la} 
Feza G\"ursey \cite{FG}.}
can be described by
\begin{eqnarray}
\vert \mbox Q \rangle\, \equiv \, 
\sum_{k,j} A_{kj}\,\vert m_k, s_j; \vec p_k \,\rangle,
\quad &&s_j= -s\hbar, (-s+1)\hbar, \ldots, (s-1)\hbar, +s\hbar\quad,
\nonumber \\
&& k= 1, 2, \ldots n\quad,
\label{a}
\end{eqnarray}
with $A_{kj}$ as some appropriately chosen superposition
coefficients.
That is, a general
quantum test particle is described by a linear superposition of
the  Casimir invariants associated with the Poincar\'e group ---
the $m_k$ are the masses, $s_j$ are the $(2 s + 1)$ spin projections (along the 
$\vec J$, for convenience) of spin $\bf s$,
and $\vec p_k$ is the momentum of the {\em kth} mass eigenstate.
 To avoid certain complicated
conceptual questions, I have refrained from summing over $s$
(which takes integral and half--integral values).

The quantum test particle as introduced here is a slight 
generalization of the particles introduced by Wigner.
Wigner's 1939 paper
suggests that quantum particles are to be specified
by the Casimir invariants associated with the 
Poincar\'e group. 
These are the $\vert m_k, s_j; \vec p_k \,\rangle$ that appear in Eq.
(\ref{a}).
The notion of a quantum test particle as presented
in Eq. (\ref{a}) is a generalization based on the 
existence of
empirically observed
particles that are a linear superposition of mass eigenstates.
The neutral $K^0$--$\overline{K}^0$ mesons, and the weak--flavor eigenstates
of neutrinos as suggested by various data \cite{vd}, fall in  a
subclass with, 
$\sum_{k,j} \rightarrow \sum_{k}$,
in Eq. (\ref{a}). With the existing laser technology,
atomic systems can be easily constructed in a state
similar to that given by Eq. (\ref{a}),
with $m_k$ replaced by $E_k$ ($E_k$ being an atomic state).
For the sake of simplicity, we will set ${\vec p_k}= {\vec 0}$.

Definition of a general quantum test particle via
Eq. (\ref{a}) allows us to introduce  two class of ``flavors;'' 
the first class for which the sum in Eq. (\ref{a}) involves a
{\em single} spin--projection independent of $m_k$. For the second
class, the sum in Eq. (\ref{a}) must contain at least {\em two
distinct} spin--projections. Eqs. (\ref{Q1}) and (\ref{Q2}), below,
provide an example of the first and second class, respectively.

It will be seen that
by making  appropriate ``flavor measurements'' one can use the
flavor states to make flavor--oscillation clocks.
We will show that flavor states of both classes
redshift identically in the gravitational 
field of a spherically symmetric mass --- i.e.,
as expected on the basis of the geometric interpretation
of general relativity.
However, when these 
identically redshifting clocks are introduced in the 
gravitational field of  a rotating source,
a splitting in the redshift of the
flavor--oscillation clocks of the second class
takes place with respect to the 
flavor--oscillation clocks of the first class.

We will consider the flavor states to be at
rest at a fixed position in the
gravitational environment.\footnote{See, for example, 
 Ref. \cite[Sec. 3.6]{OR} for the usual operational procedure
for such a general relativistic 
setup.
In particular, note \cite[pp. 166,167]{OR}:
\begin{quote}
 ``Since the behavior of freely falling clocks is completely predictable 
from the principle of equivalence, we will use freely falling clocks for all
our measurements in the gravitational field, even measurements
at a fixed position, for instance, a measurement at a fixed position on the 
surface of the Earth. For this purpose, we use a freely falling clock, 
instantaneously at rest at the fixed position. As soon as the clock 
has fallen too far from our fixed position and acquired  too much 
speed, we must replace it by a new clock,
instantaneously at rest. Whenever we speak of the time as measured by
``a clock located at a fixed position'' in a gravitational field,
this phrase  must be understood as shorthand for a complicated
measurement procedure, involving many freely falling, disposable clocks, 
used in succession.''
\end{quote}
} 
Therefore, for the situation
under consideration one need not worry about whether the particles
are described by Klein--Gordon equation, or Dirac equation, or an equation
for some higher spin \cite{hs}.\footnote{The essential elements 
of the structure that
are needed are not the vector, nor the 
spinor fields, but their spin--independent
time evolution as $\exp[\pm i\,p_\mu x^\mu]$.} It will suffice to
know that each mass eigenstate has $(2s+1)$ spinorial degrees of freedom, and
that each of these
degrees of freedom evolves in time
as 
\begin{equation}
\vert m_k,s_j\rangle \rightarrow
\exp\left[-\frac{i H t}{\hbar}\right] \vert m_k,s_j\rangle\quad.
\end{equation}
The redshift of the flavor--oscillation clocks is determined by the
gravitationally induced 
relative phases between various mass eigenstates.
Each of the mass eigenstates,
$\vert m_k,s_j\rangle$, picks up a 
gravitationally induced phase. This phase, in general, 
depends on $m_k$, and the relative orientation of $s_j$ 
with respect to $\vec J$ and
$\vec r$ ($\vec r\,\, =$
position of the test paricle).
Various mass eigenstates develop relative phases as the
quantum test particle evolves in a given gravitational 
environment. {\em These relative phases depend not only on the
gravitational source but also on the specific 
quantum mechanical characteristics of the quantum test particle
as contained in $A_{kj}$. This introduces the essential
{\em non-geometric} element when gravitational and quantum
phenomena are considered simultaneously.} Something quite close
to this was already realized by Sakurai \cite[p. 129]{JJS} when, in the
context of the 
celebrated Colella, Overhauser, and Werner experiment on
 neutron interferometry \cite{COW}
he wrote ``This experiment also shows
that gravity is not purely geometric at the quantum mechanical level because
the effect depends on $(m/\hbar)^2$,'' but noting immediately, ``However,
this does not imply that the equivalence principle is unimportant
in understanding 
an effect of this sort. If the gravitational mass ($m_{grav}$) and inertial
mass ($m_{inert}$) were unequal, $(m/\hbar)^2$ would have to be
replaced by $ m_{grav}\,m_{inert}/\hbar^2$.'' 

We take the following as working definitions.
{\em Geometrical elements} are those that are completely specified by
the gravitational source. {\em Non-geometrical} elements are those
that crucially depend on the specific details of quantum test particles
and do not follow from general relativity alone. 

In a weak gravitational field of mass,\footnote{That is, we keep
terms that are of first order in the dimensionless parameter $-GM/c^2\,r$.}
 $M$, with
spin angular momentum $\vec J(=J\hat z,\,\,\mbox{for convenience})$, 
the evolution of the mass  eigenstates is governed by the
Hamiltonian \cite{LT}
\begin{equation}
H= {\bf m} c^2 - \frac{\hbar^2}{2 m}\, 
\nabla^2   -\,\frac{G {M} {\bf m}} {r}  - 
\left(\frac{{\vec{\bf s}}}{2}\right)\cdot \vec b\quad,\label{h}
\end{equation}
and the non-relativistic Schr\"odinger equation
(recall that we shall set $\vec p = \vec 0$ for all mass eigenstates).
In Eq.  (\ref{h}), 
the gravitomagnetic field $\vec b$ is given by \cite[Eq. 6.1.25]{CW}
\begin{equation}
\vec b \equiv\frac{2 G}{c^2} \left[
\frac{ {\vec J} - 3({\vec J}\cdot\hat { r})\hat { r} }{r^3}\right]
\quad,
\end{equation}
and $\vec {\bf s}= {\bf s_z}\,\hat z$.
The quantum mechanical  operators that appear in Eq. (\ref{h}) are
defined as follows:
${\bf m} \vert m_k,s_j\rangle = m_k \vert m_k,s_j\rangle$ and
${{\bf s}}_z \vert m_k,s_j\rangle = s_j \vert m_k,s_j\rangle$.\footnote{We 
follow the notation 
that boldface letters represent quantum mechanical operators.}
$M$ and $\vec J$ will be treated as classical gravitational sources.
This framework is a  natural extension of arguments 
that were first 
put forward by 
Overhauser, and Colella \cite{OC}, and Sakurai 
\cite[pp. 126-129]{JJS}.

Consider the simplest case
with two distinct mass eigenstates of spin one half.
That is, set $n=2$ and $s=1/2$. Further,
introduce two  sets of ``flavor'' states,
one set where both the mass eigenstates have $s_j$ in the same
relative orientation, and the other set where $s_j$ are oriented in
opposite directions. First set:
\begin{equation}
 \vert Q_a \rangle \equiv \cos \theta \vert m_1, \uparrow \rangle +
\sin \theta \vert m_2, \uparrow \rangle\,, \,\,
 \vert Q_b \rangle \equiv -\sin  \theta \vert m_1, \uparrow \rangle +
\cos \theta \vert m_2, \uparrow \rangle \,\,.\label{Q1}
\end{equation}
Second set:
\begin{equation}
\vert Q_A \rangle \equiv \cos \theta \vert m_1, \uparrow \rangle +
\sin \theta \vert m_2, \downarrow \rangle\,,\,\,
\vert Q_B \rangle \equiv -\sin  \theta \vert m_1, \uparrow \rangle +
\cos \theta \vert m_2, \downarrow \rangle \,\,.\label{Q2}
\end{equation}

It should be emphasized that 
$
m_1 \ne m_2.
$ In addition, without a loss of generality, 
we take
$m_2 > m_1$.
For convenience and simplicity of the arguments, we have set
 $\vec p_k = \vec 0$; $\uparrow$ indicates $s_j = + \hbar/2$, and 
$\downarrow$ represents $s_j = - \hbar/2$.

To arrive at the stated result we now proceed in three steps.

{\bf I.} In the absence of gravity, 
let us, at time $t=0$,
prepare a system in state $\vert Q_a\rangle$.  
The flavor--oscillation probability  at a later 
time $t$ that the system is found in
state $\vert Q_b\rangle$ is:
\begin{eqnarray}
 &&P_{a\rightarrow b}(t) \nonumber\\ &&\quad= \left\vert
\langle Q_b \vert  \left\{
\exp\left[-\frac{i m_1 c^2 t}{\hbar}\right]
\cos \theta \vert m_1, \uparrow \rangle +
\exp\left[-\frac{i m_2 c^2 t}{\hbar}\right]
\sin \theta \vert m_2, \uparrow \rangle
\right\}\right\vert^{\,2} \nonumber\\
&& \quad =
\sin^2\left(2\theta\right)\,\sin^2\left[
\varphi^0\right]\quad,\label{fo}
\end{eqnarray}
where the kinematically induced phase, $\varphi^0$, is
\begin{equation}
\varphi^0\equiv \frac{(m_2-m_1) c^2 t}{2\hbar}\quad.\label{d}
\end{equation}

The similarly defined probability of flavor oscillation
for $\vert Q_A\rangle\,\rightarrow\,\vert Q_B\rangle$ is the same as above:
\begin{equation}
P_{A\rightarrow B}(t) = P_{a\rightarrow b}(t)\quad.\label{pab}
\end{equation}

The characteristic time of  flavor--oscillations,
$ a \rightleftharpoons b$ and
$A \rightleftharpoons B$, is
\begin{equation}
T^0=\frac{2\hbar}{(m_2-m_1)c^2}\quad.
\end{equation}

Thus, the phenomenon of the flavor--oscillation provides a quantum 
mechanical clock. 
In the absence of gravity, the 
flavor--oscillation clocks, $\{ a \rightleftharpoons b,\,\,A 
\rightleftharpoons B\}$,
are characterized by the same
characteristic time of  flavor--oscillations.\footnote{
Without the requirement $m_1\ne m_2$,
$\varphi^0$ would identically vanish and no flavor--oscillation 
clock shall exist. The flavor--oscillation clocks have  no
classical counterpart. If $m_1=m_2$, 
then the nearest classical counterpart
is  a gyroscope. For instance with
$m_1=m_2=m$ and $\theta=\pi/4$, 
$\vert Q_A\rangle$ becomes $\vert m,\rightarrow\rangle$ 
and 
$\vert Q_A\rangle$ becomes $\vert m,\leftarrow\rangle$.
These are the spin polarized states along the positive and negative
x-direction.}

{\bf II.}
Next, we study
the test particle evolution in the vicinity of
a  non--rotating source of gravity.\footnote{ 
We shall assume that various parameters are so 
chosen that the time scale, 
$T\left(\vert m_k,\uparrow\rangle\rightleftharpoons
\vert m_k,\downarrow\rangle\right) $, associated 
with the gravitationally induced transitions,
  is much larger
compared with the 
characteristic time of  flavor--oscillations, i.e.,
$
T\left(\vert m_k,\uparrow\rangle \rightleftharpoons
\vert m_k,\downarrow\rangle\right)  \gg T^0
$,
and that clocks can be discarded and replaced with the new ones
following a simple extension of the operational procedure \cite{OR}
outlined in footnote 3. I am 
thankful to Dr. A. Mondrag\'on (IFUNAM)
for a remark on this matter.}

The above defined probabilities are now modified by the 
gravitationally induced relative phases (each of the $m_k$ picks
up a {\em different} phase from the gravitational field):
\begin{eqnarray}
&&P_{a\rightarrow b}(t) = 
P_{A\rightarrow B}(t)  
\nonumber\\ &&\quad
= \left\vert
\langle Q_b \vert  \left\{
\exp\left[-i\varphi_1 \right]
\cos \theta \vert m_1, \uparrow \rangle +
\exp\left[-i\varphi_2\right]
\sin \theta \vert m_2, \uparrow \rangle
\right\}\right\vert^{\,2} \quad,
\end{eqnarray}
where
\begin{eqnarray}
\varphi_1
=  \left( 
m_1 c^2 - \frac{GM m_1}{r} - \frac{\hbar \hat z}{4}\cdot\vec b
\right)\frac{t}{\hbar},\,\,
\varphi_2
=  \left( 
m_2 c^2 - \frac{GM m_2}{r} - \frac{\hbar \hat z}{4}\cdot\vec b
\right)\frac{t}{\hbar}\,\,.\nonumber\\
\end{eqnarray}
The  gravitationally induced phases in $\varphi_1$ and
$\varphi_2$ that arise from the
$\vec{\bf s}\cdot\vec b$ part of $H$, Eq. (\ref{h}),
 are identical. Therefore, they  do not
contribute to the redshift--giving relative phase. The only contribution
to the redshift--giving relative phase comes from the part of the
phases that are
proportional to
$[GM m_k/r]t/\hbar$; $k=1,2$. 
A straightforward calculation yields:
\begin{equation}
P_{a\rightarrow b}(t) = 
P_{A\rightarrow B}(t) = \sin^2\left(2\theta\right)\,\sin^2\left[
\varphi^0-\varphi^M\right]\quad.
\end{equation}
The gravitationally induced phase, $\varphi^M$, reads:
\begin{equation}
\varphi^M \equiv \frac{GM}{c^2 r}\,\varphi^0\quad. \label{gm}
\end{equation}
In $\varphi^M$,
$
{GM}/{c^2 r}
$
is  the dimensionless gravitational potential,
$- \Phi$, due to a gravitational source of mass $M$. The flavor--oscillation
clocks $\{a \rightleftharpoons b,\,
A \rightleftharpoons B\}$ redshift, and redshift identically,
as expected in the
geometric interpretation of general relativity. 
The phase $\varphi^M$ is similar to the one first considered by
Good for the $K^0$--$\overline{K}^0$ mesons \cite{Good}
and studied in greater detail by Aronson, Bock, Cheng, and Fischbach
\cite{ABCF}, and Goldman, Nieto, and  Sandberg \cite{GNS}.\footnote{However,
based on the general considerations of Grossman and Lipkin
(where they consider neutrino oscillations), note should be made that 
the sign of the phase is determined by  whether one is
considering oscillations in time or distance \cite{GL}.}
The gravitationally induced neutrino--oscillation phase
given in Eq. (12) of Ref. \cite{GRF96} is a generalization
of $\varphi^M$ to the relativistic case. However the 
gravitationally induced
fractional change in the kinematic phase $\varphi^0$, 
$- \,\varphi^M/\varphi^0$, is still found to be the same, i.e., equal to
$- \,{GM}/{c^2 r}$, for both the relativistic and non-relativistic cases.
This equality assures that in the environment of a 
spherically symmetric non--rotating gravitational source,
flavor--oscillation clocks,
$\{a \rightleftharpoons b,\,
A \rightleftharpoons B\}$, redshift identically.

{\bf III.}
Finally, let us consider
the test particle evolution in the vicinity of
a  rotating source of gravity.

The above defined probabilities are now modified by the 
gravitationally induced relative phases, and these phases depend
not only on $m_k$ but also on the $s_j$ structure of the 
test particle (as contained in $A_{kj}$). 
In the first case, i.e.,
$a \rightleftharpoons b$
oscillation,
 the phase
due to the $\vec {\bf s}\cdot \vec b$  interaction
is same for both the $m_k$ and hence does not contribute
to the flavor--oscillation probability. In the second case,
i.e., $A \rightleftharpoons B$
oscillation, 
 the phase
due to the $\vec {\bf s}\cdot \vec b$ interaction
is opposite for the  two $m_k$'s (and hence becomes relative)
and contributes
to the flavor--oscillation probability. The gravitationally modified
flavor--oscillation probabilities are obtained to be:
\begin{eqnarray}
&&P_{a\rightarrow b}(t) = \sin^2\left(2\theta\right)\,\sin^2\left[
\varphi^0-\varphi^M\right]\quad,\label{clockab1}\\
&&P_{A\rightarrow B}(t) = \sin^2\left(2\theta\right)\,\sin^2\left[
\varphi^0-\varphi^M - \varphi^J\right]\quad.\label{clockab2}
\end{eqnarray}
The new gravitationally induced phase
that appears in $A \rightleftharpoons B$ 
flavor oscillations via Eq. (\ref{clockab2}) is
\begin{equation}
\varphi^J =
\left( 2\,\frac{\vec {\bf s}_z}{2}\cdot \vec b\right)
\frac{t}{2\hbar} =
\frac{G}{c^2}\left(
\frac{J-3({\vec  J}\cdot\hat { r})(\hat { r}\cdot{\hat z})}{r^3}
\right)
\frac{t}{2}\quad.\label{gj}
\end{equation}
This new contribution, $\varphi^J$, to the gravitationally induced phases
is a natural, but
conceptually important, 
 extension of Good's original considerations.\footnote{For a detailed
discussion of the gravitational phase $\varphi^M$ and its relationship 
with the pioneering work of Colella, Overhauser, and Werner
on neutron interferometry \cite{COW}, 
and its extension to neutrino oscillations 
in astrophysical contexts, the reader is referred to Refs. \cite{GRF96,AB}.}
Comparison of Eqs. (\ref{fo}) and (\ref{pab}) with Eqs.
(\ref{clockab1}) and (\ref{clockab2}) yields the result that
the flavor--oscillation
clock $a \rightleftharpoons b$ redshifts as if $\vec J$ were absent, while
the redshift 
of the flavor--oscillation clock $A \rightleftharpoons B$ depends on both
$M$ and $\vec J$. 
If the clocks $a \rightleftharpoons b$
and $A \rightleftharpoons B$ are reintroduced in the environment
of a non--rotating source they will run in synch, i.e., identically.
This is the central result of this essay.
Conceptually, 
this situation may be considered as a rough gravitational analog
of the Zeeman effect of atomic physics.\footnote{I thank
Dr. Nu Xu (LBNL) for bringing this analogy to my attention.}

The quantum--mechanically induced
gravitational phase $\varphi^J$  does not depend on $\hbar$.
This $\hbar$ independence is a generic feature of all
interaction Hamiltonians that depend linearly on the
Planck constant. However, what is remarkable here
is that the relevant interaction
Hamiltonian that gives rise to the non--geometric element obtained here,
turns out, as a consequence  of the equality of the inertial and 
gravitational masses, to be precisely of the form that removes
$\hbar$ dependence in the redshift--splitting
phase $\varphi^J$.

We now discuss in a little greater detail the origin of the 
gravitationally induced phases.
First, consider a non-rotating gravitational
source.
For a single mass eigenstate the 
classical effects of gravitation may be considered to depend
on  a force, $\vec{F}$, while the quantum--mechanical 
effects are determined by the gravitational 
interaction energy, $H^M_{int}$.
In the weak field limit,
the interaction energy and the force are
given, respectively, by
\begin{equation}
H^M_{int} = m\,  \phi\,,\quad
\vec{F}  = - \vec{\nabla}\,H^M_{int}\quad,
\end{equation}
where the gravitational potential
 $\phi = - GM/r$.
Along an eqi--$\phi$ surface 
the $\vec{F}$ vanishes and there are no classical effects in this direction.
The constant potential along a segment of an  eqi--$\phi$ surface  
can be removed by going to an appropriately accelerated frame. 
Quantum mechanically, the mass eigenstate
picks up a global phase factor
$\exp\left[- i\,  m \,\phi\, t/\hbar \right]$. Again, there are no
physical consequences. 

If we now consider a physical state that is in 
a  linear superposition
of different mass eigenstates, then 
relative phases are induced between various mass 
eigenstates.
This happens because the phase,
$\exp\left[- i\,  m \,\phi\, t/\hbar \right]$, depends on mass, $m$,
of the mass eigenstate, and by construction each mass eigenstate carries
a different mass. 
These relative phases are observable as flavor--oscillation 
phases.\footnote{The 
observability of these phases is not for  a local observer,
but for an observer making measurements stationed at a different eqi-$\phi$
surface.}
 Specifically, 
along an eqi--$\phi$ surface the gravitational force $\vec{F}$ vanishes,
while the relative quantum mechanical phases induced in the evolution
of a linear superposition of mass eigenstates do not. 
For the 
flavor--oscillation clocks, $a \rightleftharpoons b$ and 
 $A \rightleftharpoons B$,
the general expression of the gravitationally induced
flavor--oscillation phase, $\varphi^M$, is given by Eq. (\ref{gm}).

For a rotating gravitational
source, all of the above observations still remain valid. However, in
addition, one must now consider the gravitomagnetic
interaction energy and the torque,
given, respectively, by
\begin{equation}
H_{int}^J = -\,\left(\frac{ {\vec{\bf s}}}{2}\right) \cdot \vec b\,,\quad
\vec \tau =
\left(\frac{ {\vec{\bf s}}} {2}\right) \times \vec b\quad.
\end{equation}

For the set of flavor states
$\{\vert Q_A \rangle,\vert Q_B \rangle\}$, where ${\vec{\bf s}}
= (\sigma_z \hbar/2)\hat z$
and $\vec J= J\hat z$, {\underline{if}}, to empasize the
differences between quantum and classical considerations, 
 we study the evolution at $\vec r = r \hat z$,
it is immediately clear that there is no classical effect (as far as
their precession is concerned\footnote{
Note, the force exerted on the
spin
\[
\vec F =\left(\frac{1}{2} \vec{\bf s}\cdot \nabla\right)\vec b\quad,
\]
is non-zero.}) on the 
individual spins because of the vanishing $\vec \tau$,
whereas quantum mechanically the flavor--oscillation
evolution is determined by 
$H^J_{int}$--dependent phases, and these phases are  
non-zero (and opposite for
the two spin configurations superimposed in the set
$\{\vert Q_A \rangle,\vert Q_B \rangle\}$). 
The general expression for the gravitationally induced
flavor--oscillation phase $\varphi^J$ is given by Eq. (\ref{gj}).\footnote{
For a detailed discussion of gravitational redshift in the 
presence of rotation, though confined to classical test particles, 
the reader is referred to Ciufolini and Wheeler \cite{CW}.}

In this essay we presented 
a generalized notion of flavor--oscillation clocks.
The generalization contains the element that various
superimposed mass eigenstates  may have 
different relative orientation of the component of their spin 
with respect to the rotational axis of the the gravitational source.
It is found that these quantum mechanical clocks do not always 
redshift identically when moved from the gravitational
environment of a non--rotating source to the field of a rotating
source.
The gravitationally induced 
non--geometric
phase $\varphi^J$ is independent of $\hbar$ despite
its origins in  quantum  mechanical phases. Finally,
by interchanging the spin--projections associated 
with $m_1$ and $m_2$ in the 
the states $\vert Q_A\rangle$ and $\vert Q_B\rangle$ we 
 introduce a {\em third} set of flavors:
\begin{eqnarray}
\vert Q_{A^\prime} \rangle \equiv \cos \theta \vert m_1, \downarrow
\rangle +
\sin \theta \vert m_2, \uparrow \rangle\,,\,\,
\vert Q_{B^\prime} \rangle \equiv -\sin  \theta \vert m_1, \downarrow \rangle +
\cos \theta \vert m_2, \uparrow \rangle \,\,.\label{Q3}\nonumber\\
\end{eqnarray}
This results in replacing Eqs.  (\ref{clockab1}) and
(\ref{clockab2}) by:
\begin{eqnarray}
&&P_{a\rightarrow b}(t) = \sin^2\left(2\theta\right)\,\sin^2\left[
\varphi^0-\varphi^M\right]\quad,\label{clockab1p}\\
&&P_{A\rightarrow B}(t) = \sin^2\left(2\theta\right)\,\sin^2\left[
\varphi^0-\varphi^M - \varphi^J\right]\quad,\label{clockab2p}\\
&&P_{A^\prime\rightarrow B^\prime}(t) 
= \sin^2\left(2\theta\right)\,\sin^2\left[
\varphi^0-\varphi^M + \varphi^J\right]\quad.\label{clockab3p}
\end{eqnarray}
The flavor--oscillation clocks
$\{ a \rightleftharpoons b, \,A \rightleftharpoons B,\,
A^\prime \rightleftharpoons B^\prime\}$ form the maximal set of flavors
for spin one half.\footnote{The reversing of
spin--projections in states $\vert Q_a\rangle$ and $\vert Q_b\rangle$
may be used to include a fourth flavor. Such an addition to the
maximal set does not alter our conclusions.}
The sign of the gravitational phase induced by  the 
$\vec{\bf s}\cdot\vec J$ term in the Hamiltonian
carries   {\em opposite} 
signs for the
 $\,A \rightleftharpoons B$ and 
$A^\prime \rightleftharpoons B^\prime$ flavor--oscillations.
If one agrees to measure redshift average with respect
to an equally weighted ensemble of the three flavor types, then
the non--geometric element averages out to zero. Extension to higher spins
being obvious, we propose that the non--geometric element in redshifts
may be interpreted as a quantum mechanically induced fluctuation over  a
geometric structure of space--time. In the weak field limit, the 
amplitude of these fluctuations is directly proportional to the product of
the rotation, as measured by $\vec J$, and the spin of the test particle.

{\bf Acknowledgements} 
The author is grateful to Drs. Mikkel Johnson
(LANL) and Nu Xu (LBNL)
for their comments, and ensuing  discussions,  on 
several drafts of this manuscript.

\end{document}